\documentclass[twocolumn,showpacs,preprintnumbers,superscriptaddress,prl,amsmath,amssymb,nofootinbib]{revtex4-1}

\usepackage{graphicx}
\usepackage{dcolumn}
\usepackage{bm}
\usepackage{color}
\usepackage{ulem}
\usepackage{gensymb}
\usepackage{braket}
\usepackage{epstopdf}
\usepackage{amsmath}
\usepackage[percent]{overpic}

\newcommand{\MBT}{\,MnBi$_2$Te$_4$}
\newcommand{\MST}{\,MnSb$_2$Te$_4$}

\begin{document}

\title{The elusive quantum anomalous Hall effect in \MBT: a materials perspective}

\author{J.-Q. Yan}
\email{yanj@ornl.gov}
\affiliation{Materials Science and Technology Division, Oak Ridge National Laboratory, Oak Ridge, Tennessee 37831, USA}

\date{\today}

\begin{abstract}
Observation of the quantum anomalous Hall effect (QAHE) in \MBT~flakes is one of the most exciting results in the study of the intrinsic magnetic topological insulator \MBT~and related compounds. This fascinating result is yet to be reproduced two years after its first report. The quality of starting  \MBT~single crystals is believed to be the key factor. An interesting and important question to address is what is the right quality to enable the QAHE. In this perspective, we present possible approaches to tuning the magnetic and topological properties of \MBT~by using lattice imperfections, strain, stacking sequence, and interactions between the substrate and flakes/films. It is of critical importance to eventually identify the factor(s) responsible for the realization of QAHE.

\end{abstract}

\maketitle
\MBT~ attracted intense attention in the last few years after it was investigated as the first intrinsic antiferromagnetic topological insulator\cite{otrokov2019prediction,gong2019experimental}. As shown in Fig. 1, \MBT~consists of van der Waals bonded septuple layers along the crystallographic $c$-axis. Each septuple layer (SL) can be viewed as inserting a MnTe bilayer into a quintuple layer. This beautiful natural heterostructure provides an intimate and natural combination of magnetism and electronic band topology, providing a new materials platform for the study of various exotic phenomena. Below T$_N\simeq$\,25\,K, magnetic moments on Mn ions order ferromagnetically in each SL with moments along the $c$-axis; an antiferromagnetic interlayer coupling leads to the so-called A-type antiferromagnetic order. The competing antiferromagnetic and ferromagnetic interactions in each SL and the interlayer antiferromagnetic coupling can be understood following the famous Goodenough-Kanamorri rules that were formulated first by Prof. John B. Goodenough and mathematically demonstrated by  Prof. Junjiro Kanamori in the 1950s\cite{goodenough1955theory, goodenough1958interpretation, kanamori1959superexchange}.  The resulting A-type magnetic order has important consequences on the magnetic and topological properties of \MBT~especially in the atomic limit. The magnetic order breaks time-reversal symmetry $\Theta$. However, $\Theta$T$_{1/2}$, where T$_{1/2}$ is the translation vector between oppositely aligned Mn layers, preserves the topological invariant and enables an antiferromagnetic topological insulator\cite{mong2010antiferromagnetic}. This A-type antiferromagnetism also leads to thickness dependent magnetic and topological properties. Flakes/films with an odd(even) number of SLs have uncompensated(compensated) magnetism that makes possible the experimental realization of various quantum phenomena including quantum anomalous Hall effect (QAHE)\citep{deng2020quantum} and axion insulator\cite{liu2020robust}. The observation of QAHE in a 5-SL device agrees well with the theoretical prediction\citep{deng2020quantum}. However, this important result has not been reproduced two years after its first report. The extreme challenge of reproducing indicates materials-related issues that should be resolved in order to advance this field.  There are other controversial results that also point to the materials-related issues. For example,
the size of the magnetism-induced gap in topological surface states of \MBT~ is under hot debate, and a zero plateau quantum anomalous Hall state was observed in both 5-SL\citep{ovchinnikov2021intertwined} and 6-SL \MBT~ flakes\cite{liu2020robust}.

\begin{figure*} \centering \includegraphics [width = 0.80\textwidth] {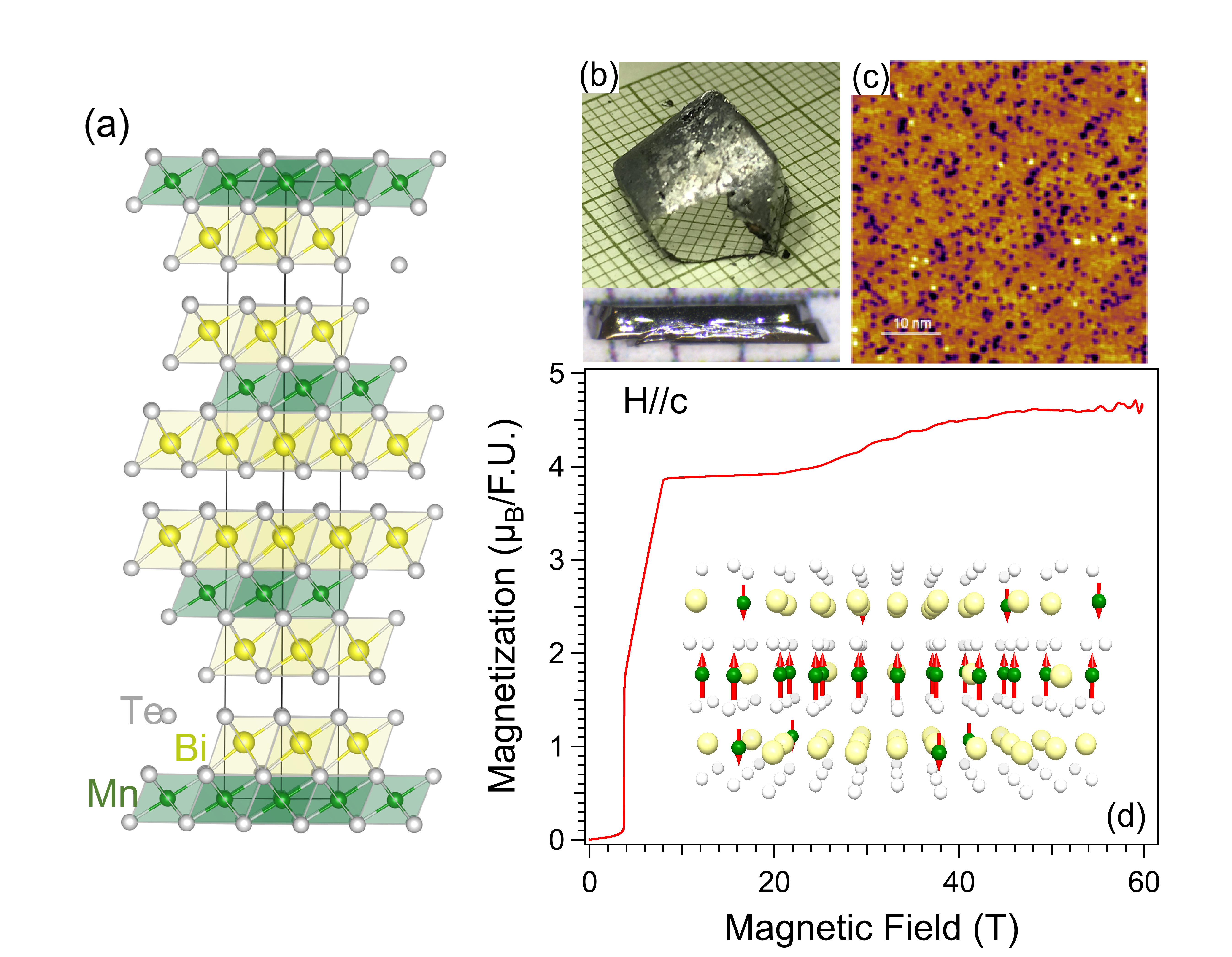}
\caption{(color online) (a) Crystal structure of \MBT. (b) \MBT~crystals grown by flux (upper) and vapor transport(lower panel) techniques. (c) STM images showing the presence of two types of defects:  bright circular protrusions (Bi$_{Te}$) and dark clover-shape depressions (Mn$_{Bi}$). (d) The field dependence of magnetization of \MBT~in magnetic fields up to 60\,T applied along the crystallographic $c$-axis. The 1st plateau occurs around 10\,T with the magnetization about 3.9$\mu_B$ due to the ferrimagnetic arrangement of Mn moments. The 2nd plateau is observed around 50\,T with the magnetization of about 4.6$\mu_B$ after all Mn moments are polarized. The magnetization data are replotted from Ref.[\citenum{lai2021defect}]. The inset illustrates the spin alignment in a ferrimagnetic septuple layer using the same color scheme for the elements as panel (a). }
\label{fig1-1}
\end{figure*}

\section{Current Status}

\MBT, like other crystalline solids, has defects in its crystal lattice. The population and distribution of lattice defects depend on the synthesis approaches and detailed processing parameters. In flux-grown and vapor-transported \MBT~crystals synthesized in our lab, scanning tunneling microscopy(STM) found 2-4\% Mn$_{Bi}$ (Mn at Bi site), 4-15\% Bi$_{Mn}$ (Bi at Mn site), and about 0.2\% Bi$_{Te}$ antisite defects (see Fig. 1). In our notation, 1\% of Mn$_{Bi}$ indicates 2\% Bi$_{Mn}$ in Mn stoichiometric or deficient \MBT~ because Bi$_{Mn}$ has a much lower formation energy than V$_{Mn}$ (Mn vacancy)\cite{du2021tuning}.

These lattice defects can account for the metallic conducting behavior observed experimentally in as-grown crystals. Those magnetic defects complicate the initially believed A-type antiferromagnetism in \MBT.  A careful neutron single crystal diffraction study confirmed the antiferromagnetic coupling of the Mn$_{Bi}$ ions to Mn$_{Mn}$, forming ferrimagnetic SLs in \MBT. This ferrimagnetic arrangement of Mn moments reduces the net magnetization of each SL. This makes magnetization measurement an effective probe of the population of antisite magnetic defects. The antiferromagnetic coupling between Mn$_{Bi}$ and Mn$_{Mn}$ can also be understood following Goodenough-Kanamorri rules. This interaction is very strong and a large magnetic field of approximately 50\,T is needed to polarize all Mn moments (see Fig. 1)\citep{lai2021defect}. Increasing the amount of Mn$_{Bi}$  could be a valid approach to inducing ferromagnetic interlayer interactions in \MBT, which would be beneficial for QAHE. This is well illustrated in \MST~  where the concentration of magnetic defects can be tuned in a wider range due to the similarity of ionic size and electronegativity between Mn and Sb\citep{liu2021site}. An antiferromagnetic interlayer coupling in \MST~ for a low concentration of Mn$_{Sb}$ gives way to a ferromagnetic interlayer interaction with increasing Mn$_{Sb}$.

The random distribution of Mn$_{Bi}$ and Bi$_{Mn}$ antisites results in spatial fluctuation of the local density of states (LDOS) near the Fermi level\cite{huang2020native}. Scanning tunneling spectroscopy (STS) found clustering of Mn$_{Bi}$ antisites significantly suppresses LDOS at the conduction band edge, and Bi$_{Mn}$ possesses a localized electronic state 80\,meV below E$_F$ contributing to a pronounced peak in the LDOS. Currently available \MBT~ crystals all have a high density of lattice defects preventing a careful  investigation of the intrinsic properties. Interestingly, quantum phenomena were observed despite those lattice defects and resulting electronic inhomogeneities. Recently, lattice disorder in \MBT~was proposed to induce the quantum Hall edge state coexisting with the QAH edge state in magnetic fields\cite{li2021coexisting} and to modulate the gap size of the surface states\cite{garnica2021native, shikin2021sample, sitnicka2021systemic}.

Chemical doping has been employed to effectively tune the magnetic and topological properties of \MBT.  As in transition metal doped Bi$_2$Te$_3$, partial substitution of Bi by Sb in \MBT~ can tune the Fermi level and induces a crossover from n-type to p-type conducting behavior\cite{yan2019evolution,chen2019intrinsic}. However, Sb substitution reduces the spin orbital coupling and increases the population of magnetic antisite defects. Both are detrimental to the band inversion necessary for electronic topology. Partial substitution of Mn by Sn dilutes the magnetic sublattice which suppresses T$_N$ and the critical magnetic field for the spin flop transition\cite{zhu2021magnetic}. Ge and Pb substitutions are expected to play a similar role. Vapor transport\cite{yan2021vapor} and solid state syntheses of Cr- and V-doped \MBT~ efforts were in vain.

High pressure is a clean means to tune the magnetic properties and electronic band structure of quantum materials. Under hydrostatic pressure up to 12.5GPa, T$_N$ of \MBT~first increases slightly with pressure and then decreases until 7GPa above which the transport measurement indicates the absence of long range magnetic order\cite{chen2019suppression}. Hydrostatic pressure modifies both the interlayer and intralayer magnetic interactions; its influence on the topological properties is to be revealed. Pressure studies of crystals with controlled defect chemistries and concentrations may provide further insights.

\section{Future Needs and Prospects}

Although remarkable progress has been made in the synthesis of \MBT~ crystals and films, and in the property tuning via chemical doping, pressure, magnetic/electric fields, it is still not clear whether and how the \MBT~ flakes used in Ref. [\citenum{deng2020quantum}] are different from others. Defect engineering, stacking and strain field effects, and interfacial engineering may provide the key information to resolve the puzzle.

\textit{Defect engineering}: The presence and abundance of lattice defects make possible the fine tuning of the magnetism and the electronic band structure of \MBT~and related compounds by controlling the type, concentration, and distribution of lattice defects. At this time, it is known that \MBT~crystals with the least or no Bi$_{Mn}$ facilitates the observation of Chern insulator state under high magnetic fields\cite{yan2021vapor}. The transport results in Ref. [\citenum{deng2020quantum}] suggest those \MBT~ crystals used in that work should be nearly Mn stoichiometric. The amount and distribution of Mn$_{Bi}$ might distinguish their crystals from others. Increasing the amount of Mn$_{Bi}$ enhances the ferromagnetic interlayer interaction and might eventually lead to  ferromagnetic \MBT~at the expense of spin orbital coupling.  Reducing the defect concentration to be minimal is essential to explore the intrinsic physical properties of \MBT. It should be mentioned that the energy difference between the ferromagnetic and antiferromagnetic ground states is small in \MBT, which makes possible a temperature-induced ground state change at a certain critical concentration of Mn$_{Bi}$. This seems to provide a reasonable explanation for the antiferromagetism at high temperatures and the QAHE at low temperatures reported in Ref. [\citenum{deng2020quantum}]. If this is the case, however, the low-temperature ferromagnetism is irrelevant to the thickness dependent magnetism in \MBT~flakes.

Currently available \MBT~crystals grown out of Bi$_2$Te$_3$ flux\cite{yan2019crystal} and those by chemical vapor transport technique\cite{yan2021vapor,hu2021growth}  have comparable amount of Mn$_{Bi}$ because in both syntheses the crystallization occurs in a narrow temperature window of 500-600$^\circ$C. This indicates the importance of thermodynamics over the kinetics for the formation of Mn$_{Bi}$ in this temperature range. Further advancements in crystal growth are needed to vary the growth temperature in a wider range or the growth mechanism thus to control the population and distribution of Mn$_{Bi}$.  Post-growth thermal treatment of currently available crystals provides a different but maybe easier route to defect engineering of \MBT~ and  related compounds.

\begin{figure} \centering \includegraphics [width = 0.47\textwidth] {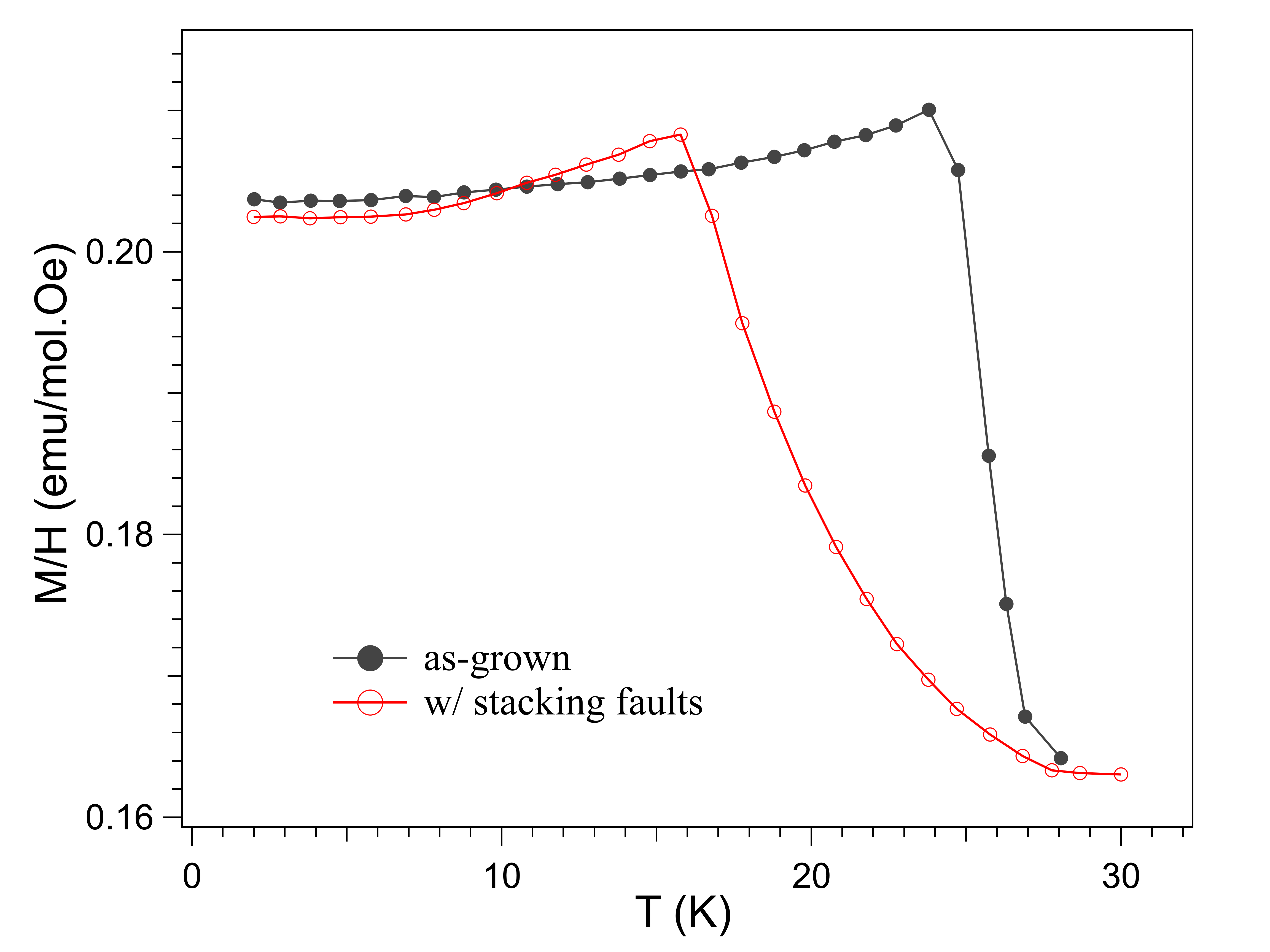}
\caption{(color online) Temperature dependence of magnetization of the same piece of \MBT~crystal before and after bending back and forth several times. This deformation is expected to introduce stacking faults and some strain field. The suppressed T$_N$ changes back to the original value in about one hour. This is different from that seen in RuCl$_3$, where the new stacking configuration induced by deformation is stable\citep{cao2016low}. }
\label{Bending-1}
\end{figure}

\textit{Interfacial Engineering}: The interlayer interaction in van der Waals materials can be affected by the interlayer stacking sequence or external electric fields.  Recent work suggests that electric fields directly modify the electronic band structure of \MBT~ instead of via affecting the magnetism\citep{cai2021electric}. Stacking configuration may provide an effective and convenient control of the magnetism and hence the topological properties of \MBT. As in other van der Waals magnets, stacking faults in \MBT~may form during materials synthesis, or during mechanical exfoliation or device fabrication. There are ample examples where stacking faults modify the magnetic properties in van der Waals magnets.  For example, in RuCl$_3$, T$_N$ depends on the stacking sequence of RuCl$_3$ layers\cite{cao2016low}; in CrI$_3$, where ferromagnetism survives in monolayer, the layer stacking leads to antiferromagnetism in bilayer and different ferromagnetic states in trilayer\cite{zhang2021emergent}; in Fe$_5$GeTe$_2$, an itinerant exfoliable ferrromagnet, partial substitution of Fe by Co changes the stacking of the building blocks and thus the magnetic properties\cite{may2020tuning}. Figure\,\ref{Bending-1} shows our preliminary results that demonstrate that stacking faults suppress T$_N$ and this indicates that controlling the lateral stacking might eventually lead to a ferromagnetic interlayer coupling that would facilitate the observation of QAHE. More theoretical and experimental efforts are needed to precisely manipulate the lateral sliding of SLs and to understand the corresponding modification of  magnetic and topological properties. Twisting the SLs provides a different approach to manipulate the stacking. A recent theoretical study on twisted bi-SL \MBT~moire superlattice suggests that the twisted stacking facilitates the emergence of QAHE in addition to other possible correlated topological phases\cite{lian2020flat}.

The heterointerface between \MBT~flakes and substrate or capping layer may also play a role. Thin \MBT~flakes are usually prepared by micro-mechanical exfoliation and then transferred on a SiO$_2$/Si substrate. The interaction between the substrate and \MBT~flakes has not been investigated so far. This interfacial interaction might extend into the flake and affect the interactions between SLs. Raman spectroscopy studies observed an intense ultralow frequency peak in 2-4 QL Bi$_2$Te$_3$ which is attributed to a substrate induced interface mode\cite{zhao2014interlayer}. One would naturally ask whether similar interfacial effects exist between the SiO$_2$/Si substrate and \MBT~flakes/film or between flakes and the encapsulation layer. The effects of interfacial interactions on the magnetic and electronic properties are of particular interest.

\textit{Strain engineering}: The above mentioned deformation process and the heterointerface interaction between flakes/films and substrate may also introduce some strain fields to \MBT. The effects of different types of in-plane and out-of plane strain fields on the electronic and magnetic properties of \MBT~ crystals, films, or flakes deserve some efforts and  might lead to new physics and novel exotic phenomena. Related to the in-plane strain fields is the longitudinal displacement along the $c$-axis, which can be efficiently adjusted using uniaxial pressure  due to the weak van der Waals coupling between the layers.

\section{Conclusions}

In summary, \MBT, as the first intrinsic magnetic topological insulator, provides an excellent materials playground for novel quantum phenomena. Efforts on fine tuning of the magnetism and electronic band topology via manipulation of defects, stacking, strain field, and interfacial interactions may eventually identify the factors responsible for QAHE observed in Ref. [\citenum{deng2020quantum}]. Density functional theory calculations that could include the diverse distribution of magnetic defects in addition to on-site Coulomb interactions are needed in order to understand the close interplay between magnetism and electronic band topology in \MBT~ with rich defects.

\section{Acknowledgment}
The author thanks Matthew Brahlek, Cuizu Chang,  Maohua Du, Andrew May, Michael McGuire, Robert McQueeney, Hu Miao, Ni Ni, Brian Sales,  Linlin Wang, Weida Wu, and Xiaodong Xu for  discussions, Zengle Huang and Weida Wu for providing the STM image, Andrew May and Michael McGuire for improving the manuscript. Work at ORNL was supported by the U.S. Department of Energy, Office of Science, Basic Energy Sciences, Materials Sciences and Engineering Division.

 This manuscript has been authored by UT-Battelle, LLC, under Contract No.
DE-AC0500OR22725 with the U.S. Department of Energy. The United States
Government retains and the publisher, by accepting the article for publication,
acknowledges that the United States Government retains a non-exclusive, paid-up,
irrevocable, world-wide license to publish or reproduce the published form of this
manuscript, or allow others to do so, for the United States Government purposes.
The Department of Energy will provide public access to these results of federally
sponsored research in accordance with the DOE Public Access Plan (http://energy.gov/
downloads/doe-public-access-plan).

\section{references}
\bibliographystyle{apsrev4-1}
%

\end{document}